\documentclass[aps,pre,twocolumn,showpacs,groupedaddress]{revtex4}

\usepackage{graphicx}
\usepackage{dcolumn}
\usepackage{bm}
\usepackage{amssymb}
\usepackage{color}

\begin{document}

\title{Dynamics of Fluid Vesicles in Oscillatory Shear Flow}
\author{Hiroshi Noguchi}
\email[]{noguchi@issp.u-tokyo.ac.jp}
\affiliation{
Institute for Solid State Physics, University of Tokyo,
 Kashiwa, Chiba 277-8581, Japan}


\begin{abstract}
The dynamics of fluid vesicles in oscillatory shear flow was studied using  
differential equations of two variables: the Taylor deformation parameter 
and inclination angle $\theta$.
In a steady shear flow with a low viscosity $\eta_{\rm {in}}$ of internal fluid,
the vesicles exhibit steady tank-treading motion
with a constant inclination angle $\theta_0$.
In the oscillatory flow with a low shear frequency,
$\theta$ oscillates between $\pm \theta_0$ or around $\theta_0$ 
for zero or finite mean shear rate $\dot\gamma_{\rm m}$, respectively.
As shear frequency $f_{\gamma}$ increases, 
the vesicle oscillation becomes delayed with respect to the shear oscillation,
and the oscillation amplitude decreases.
At high $f_{\gamma}$ with $\dot\gamma_{\rm m}=0$,
another limit-cycle oscillation between $\theta_0-\pi$ and $-\theta_0$ is found to appear.
In the steady flow,
$\theta$ periodically rotates (tumbling)
at high  $\eta_{\rm {in}}$, and 
$\theta$ and the vesicle shape oscillate (swinging) at middle  $\eta_{\rm {in}}$ and high shear rate. 
In the oscillatory flow, the coexistence of two or more limit-cycle oscillations
can occur for low $f_{\gamma}$  in these phases.
For the vesicle with a fixed shape,
the angle $\theta$ rotates back to the original position after an oscillation period.
However, it is found that a preferred angle can be induced by
small thermal fluctuations.
\end{abstract}

\maketitle

\section{Introduction}

The dynamics of soft deformable objects such as liquid droplets~\cite{rall84,ston94}, 
lipid vesicles~\cite{haas97,abka05,made06,kant05,kant06,desc09,desc09a,kant07,krau96,seif99,misb06,dank07,lebe07,lebe08,turi08,vlah07,nogu04,nogu05,nogu07b,nogu09,mess09}, 
red blood cells (RBCs)~\cite{skal90,fung04,fisc78,abka07,abka08a,kell82,tran84,naka90,wata06,pozr03,pozr05,skot07,dupi07,macm09,nogu05b,mcwh09,nogu09b,nogu09c}, 
and synthetic capsules~\cite{chan93,walt01,kess08,sui08,bagc09,kess09,lac08,lefe08} in flows has received 
growing attention experimentally, theoretically, and numerically in recent 
years.  
Under typical experimental conditions, these objects have a constant internal volume $V$ 
unless large shear breaks them.
Lipid vesicles and RBCs essentially have a constant surface area $A$,
because the number of lipids of their membranes is maintained constant in typical experimental time scales.
Under these constraints, they show rich behaviors under flows.
Flow can induce large shape deformations of these objects
and lead to transitions of dynamic modes.

\begin{figure}
\includegraphics[width=5.5cm]{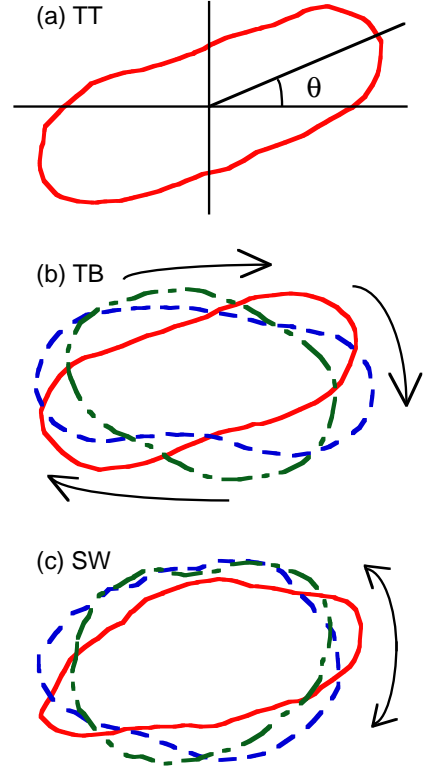}
\caption{ \label{fig:std_snap}
(Color online)
Sliced snapshots of fluid vesicles in steady shear flow
obtained by numerical simulations \cite{nogu07b} at $V^*=0.78$ and $\eta_{\rm {in}}^*=1$.
(a) Tank-treading (TT) motion at $\eta_{\rm {mb}}^*=0$ and $\dot\gamma^*=1.84$ ($\kappa/k_{\rm B}T=20$), 
where the inclination angle $\theta$ has a constant value.
(b) Tumbling (TB) motion at  $\eta_{\rm {mb}}^*=2.9$ and $\dot\gamma^*=0.92$ ($\kappa/k_{\rm B}T=40$), 
where $\theta$ rotates.
(c) Swinging (SW) motion at $\eta_{\rm {mb}}^*=2.9$ and $\dot\gamma^*=3.68$ ($\kappa/k_{\rm B}T=10$),
where $\theta$ oscillates.
The other parameters are the same as in Fig. 1 of Ref. {\cite{nogu07b}}.
}
\end{figure}

An isolated fluid vesicle exhibits three types of dynamic modes in a steady shear flow, 
with flow velocity ${\bf v}=\dot\gamma y {\bf e}_x$, 
where ${\bf e}_x$ is a unit vector in the flow direction.
When the viscosity of internal fluid $\eta_{\rm {in}}$ and membrane viscosity $\eta_{\rm {mb}}$
are low, the vesicle exhibit a tank-treading (TT) rotation with a stationary 
shape and a constant inclination angle $\theta>0$ [see Fig. \ref{fig:std_snap}(a)].
At high $\eta_{\rm {in}}$ and  $\eta_{\rm {mb}}$,
the vesicle exhibit a tumbling (TB) motion, where $\theta$ rotates [see Fig. \ref{fig:std_snap}(b)].
Around the TT-TB transition viscosity with high shear rate $\dot\gamma$,
a swinging (SW) motion appears \cite{kant06,desc09,desc09a,misb06,dank07,lebe07,lebe08,nogu07b,mess09},
where $\theta$ and the vesicle shape oscillate [see Fig. \ref{fig:std_snap}(c)].
This motion is also called trembling \cite{kant06,desc09,desc09a,lebe07,lebe08} or vacillating-breathing \cite{misb06,dank07}.
These three types of motion can be understood by
the perturbation theories for quasi-spherical vesicles
\cite{dank07,lebe07,lebe08} or a generalized 
Keller-Skalak (KS) theory for deformable ellipsoidal vesicles \cite{nogu07b}. 

RBCs and synthetic capsules also show the TT-TB transition with increasing viscosity $\eta_{\rm {in}}$.
The shear elasticity of their membranes gives
an energy barrier to the TT membrane rotation \cite{skot07}.
Since a sufficiently large shear is necessary to induce membrane rotation,
transition from the TB to TT motions occurs with increasing $\dot\gamma$ at low $\eta_{\rm {in}}$ \cite{skot07,abka07}.
The synchronization and intermittency of  $\theta$ and membrane rotations appear
at the transition region between TB and TT \cite{skot07,nogu09b}. 
At much higher $\dot\gamma$, this energy barrier becomes negligible, so
the dynamics becomes similar to that of lipid vesicles.
Thus, lipid vesicles, which have no shear elasticity, can be considered as a simple model system of RBCs in high shear rates.

The relaxation dynamics of soft objects can be examined using time-dependent flows.
However, compared to steady-flow conditions,
the dynamics in time-dependent flows has been much less investigated.
Recently, a membrane wrinkling after inversion 
of an elongational flow~\cite{kant07} and shape or orientational oscillation in structured channels
\cite{nogu09a} were discovered for fluid vesicles.
For RBCs, a shape oscillation under oscillatory shear flow was
observed experimentally~\cite{wata06} and explained by the extended KS theory \cite{nogu09c},
which gives differential equations of three variables, a shape parameter, the inclination angle $\theta$, and phase angle $\phi$.
At high shear frequency, multiple limit cycles coexist.
At middle shear amplitudes, RBCs show complex behavior such as intermittent oscillation and complicated domain boundary \cite{nogu09c}.

In this paper, we study the dynamics of a fluid vesicle in oscillatory shear flow
with  $\dot\gamma= \dot\gamma_{\rm m} +  \dot\gamma_0 \sin(2\pi f_{\gamma} t)$
using the generalized KS theory \cite{nogu07b}, which has two variables, a shape parameter and the inclination angle $\theta$.
The reduction of variables from three to two allows us to investigate the details of the dynamics. 
We investigate i) what kinds of oscillatory dynamics appear with respect to the TT, TB, and SW motions
for zero mean shear rate ($\dot\gamma_{\rm m}=0$), ii)
how the vesicle responds to the shear  oscillation with finite mean shear rate $\dot\gamma_{\rm m}$, and
iii) how the thermal fluctuations affect the dynamics.

The vesicle dynamics is described by several dimensionless quantities.
The relative ratio of the volume $V$ and surface area $A$ is characterized by
the reduced volume $V^*= 3V/(4\pi A)^{3/2} = (R_{\rm V}/R_{\rm S})^3$ or
the excess area $\Delta_{\rm S}=A/R_{\rm V}^2 -4\pi  = 4\pi\{(1/V^*)^{2/3}-1\}$,
where $R_{\rm V}= (3V/4\pi)^{1/3}$ and $R_{\rm S}=(A/4\pi)^{1/2}$.
The relative viscosity of the inside fluid and membrane are 
$\eta_{\rm {in}}^*=\eta_{\rm {in}}/\eta_0$
and $\eta_{\rm {mb}}^*=\eta_{\rm {mb}}/\eta_0 R_{\rm S}$, where 
$\eta_0$ is the viscosity of the outside fluid. 
The shape relaxation time of the vesicles with bending rigidity $\kappa$ is given
by $\tau=\eta_0 R_{\rm S}^3/\kappa$ (for $\eta_{\rm {in}}^*=1$). This
time is used to define a reduced shear rate $\dot\gamma^*=\dot\gamma \tau$.

The theory and results for the steady flow are  explained in Secs. \ref{sec:theory} and  \ref{sec:std}, respectively.
The results of the generalized KS theory are compared with those of perturbation theories, experiments, and simulations.
The dynamics for zero and finite mean shear rate is described in Sec. \ref{sec:osc} and in Sec. \ref{sec:bias}, respectively.
In Sec. \ref{sec:jef}, 
the effects of the thermal fluctuations are investigated.
Summary is given in Sec. \ref{sec:sum}.

\section{Generalized Keller-Skalak theory} \label{sec:theory}

Keller and Skalak \cite{kell82}
 analytically derived the equation of motion of vesicles or elastic capsules
based on Jeffery's theory \cite{jeff22}.
In the KS theory,
the vesicles are assumed to have a fixed ellipsoidal shape, 
\begin{eqnarray}
\Big(\frac{x_1}{a_1}\Big)^2 +\Big(\frac{x_2}{a_2}\Big)^2 +\Big(\frac{x_3}{a_3}\Big)^2 =1, 
\end{eqnarray}
where $a_i$ denote the semi-axes of the ellipsoid, and
the coordinate axes $x_i$ point along its principal directions. 
The $x_1$ and $x_2$ axes, with $a_1>a_2$, are on the vorticity ($xy$) plane,
and the $x_3$ axis is  in the vorticity ($z$) direction.
The Taylor deformation parameter is defined as $\alpha_{\rm D}=(L_1-L_2)/(L_1+L_2)=(a_1-a_2)/(a_1+a_2)$, where the maximum lengths in three directions are $L_1=2a_1$, $L_2=2a_2$, and $L_3=2a_3$.
The velocity field on the membrane is assumed to be 
\begin{eqnarray}
{\bf v}^{\rm {m}}=
               \omega \Big(-\frac{a_1}{a_2}x_2,\frac{a_2}{a_1}x_1,0\Big).
\label{eq:KS-vel}
\end{eqnarray}
The vesicle motion is derived from the energy balance 
between supply from the external fluid of the vesicle and dissipation inside the vesicle and on the membrane.
\begin{eqnarray}
\frac{d\theta}{dt}
 &=& \frac{\dot\gamma}{2}\{-1+B\cos(2\theta)\} 
\label{eq:thetb} \\
\label{eq:KS-B}
B &=& f_0\left\{f_1+ \frac{f_1^{-1}}
      {1+f_2(\eta_{\rm {in}}^* -1)
                  + f_2f_3 \eta_{\rm {mb}}^*}\right\}
\end{eqnarray}
The membrane-viscosity term was derived by Tran-Son-Tay {\it et al.} 
\cite{tran84}.
Factor $B$ is a function of the vesicle shape and viscosity 
$\eta_{\rm {in}}^*$ and $\eta_{\rm {mb}}^*$. Here,
$f_0$, $f_1$, $f_2$, and $f_3$ are given by
\begin{eqnarray*}
f_0 &=& 2/(a_1/a_2+a_2/a_1) = (1-\alpha_{\rm D}^2)/(1+\alpha_{\rm D}^2),\\
f_1 &=& 0.5(a_1/a_2-a_2/a_1) = 2\alpha_{\rm D}/(1-\alpha_{\rm D}^2),\\
f_2 &=& 0.5g(\alpha_1^2+\alpha_2^2),\\
f_3 &=& 0.5E_{\rm s}R_0/(f_1^2V),\\
g &=& \int_0^\infty (\alpha_1^2+s)^{-3/2}(\alpha_2^2+s)^{-3/2}
          (\alpha_3^2+s)^{-1/2}ds,\\
\alpha_i &=& a_i/(a_1a_2a_3)^{1/3},
\end{eqnarray*}
where $E_{\rm s}$ is an integral of shear stress over the membrane surface \cite{tran84,nogu05}.

The KS theory is extended to include the shape deformation of fluid vesicles 
 on the basis of the perturbation theory \cite{seif99,misb06,lebe07,lebe08} of quasi-spherical vesicles in Ref. {\cite{nogu07b}}.
The equation of the shape parameter $\alpha_{\rm D}$ is given by
\begin{equation}
\label{eq:ald}
\frac{d \alpha_{\rm D}}{dt} = \dot\gamma \left\{1-\left(\frac{\alpha_{\rm D}}
                   {\alpha_{\rm D}^{\rm {max}}}  \right)^2\right\}
    \left\{ -\frac{A_0}{\dot\gamma^* \kappa V^*} 
       \frac{\partial F}{\partial \alpha_{\rm D}} 
                +  A_1\sin(2\theta)\right\},
\end{equation}
where $A_0= 45/8\pi(32+23\eta_{\rm {in}}^*+16\eta_{\rm {mb}}^*)V^*$ and 
$A_1= 30/(32+23\eta_{\rm {in}}^*+16\eta_{\rm {mb}}^*)$.
Here, the terms of $\eta_{\rm {mb}}^*$ are added in $A_0$ and $A_1$
  based on the theory in Refs. {\cite{lebe07,lebe08}}. This revision
  improves the $\eta_{\rm {mb}}^*$ dependence of fluid
  vesicles in Ref. {\cite{nogu07b}}.
The first and second terms in the last parentheses represent
the forces of the bending elasticity of the membrane and the external shear stress, respectively.
The free energy is given by $F = (\kappa/2) \int (C_1+C_2)^2 dA$,
where $C_1$ and $C_2$ are the 
principal curvatures at each point of the membrane. 
Here $F$ is numerically calculated for ellipsoidal vesicles with 
$(x_1/a_1)^2 + (x_2/a_2)^2 + (x_3/a_3)^2=1$.
The prolate ($a_1>a_2=a_3$) and oblate ($a_1=a_2>a_3$) shapes are energy 
minima and maxima, respectively,
and $\partial F/\partial \alpha_{\rm D}$ diverges in the limit of maximum
extension, $\alpha_{\rm D} \to \alpha_{\rm D}^{\rm {max}}(V^*)$.
Equations~(\ref{eq:thetb}) and (\ref{eq:ald})
 are numerically integrated
using the fourth-order Runge-Kutta method with a time step $\Delta t \leq 0.0005$.

For the data in Ref. \cite{nogu07b} and in Sec. \ref{sec:std}, 
the interpolations of $F$ and prefactors in Eq.(\ref{eq:KS-B}) are employed.
For the dynamics in the oscillatory flow at $V^*=0.9$ and $\eta_{\rm {mb}}^*=0$,
the fit functions are employed instead to avoid the artifacts of non-smooth functions:
$(1/\kappa) \partial F/\partial \alpha_{\rm D}= 30 \alpha_{\rm D} -820 \alpha_{\rm D}^5 -1.7/\sqrt{\alpha_{\rm D}^{\rm {max}}-\alpha_{\rm D}} +1.7/\sqrt{\alpha_{\rm D}^{\rm {max}}}$,
$f_2= 0.3185+0.6 \alpha_{\rm D}^2 +14 \alpha_{\rm D}^6 + \exp\{124 (\alpha_{\rm D}-0.39)\}$, and
$\alpha_{\rm D}^{\rm {max}}=0.356$.
The difference in the phase boundaries calculated with the interpolation and fit functions is very small, $\Delta\eta_{\rm {in}}^* < 0.04$.

\begin{figure}
\includegraphics{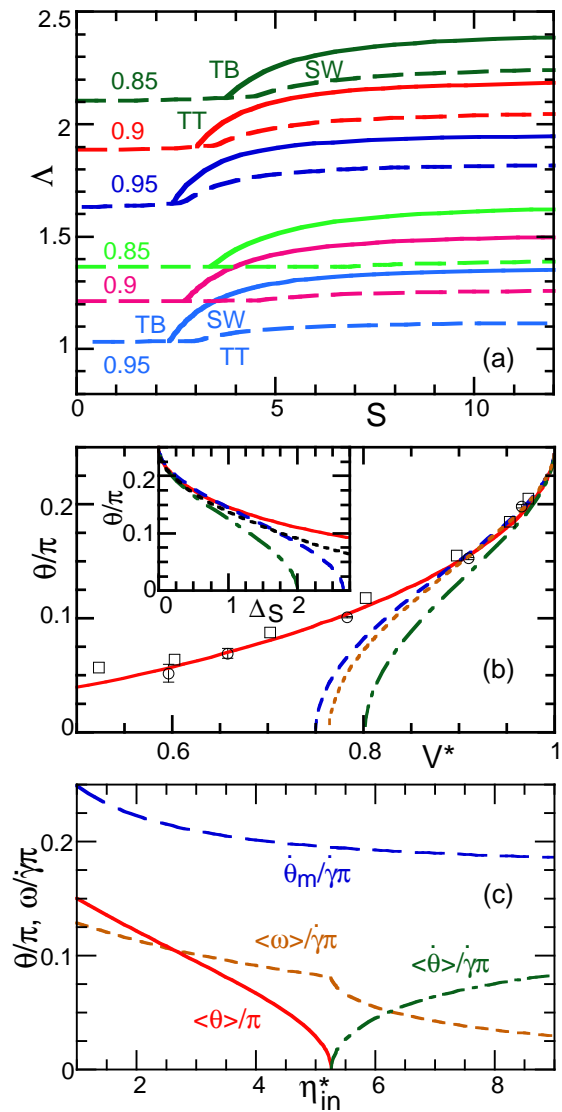}
\caption{ \label{fig:phase}
(Color online)
Dynamics of fluid vesicles in steady shear flow.
(a) Dynamical phase diagrams as a function of 
two dimensionless parameters $S$ and $\Lambda$
at $V^*=0.85, 0.9$, and $0.95$.
The dashed lines represent the boundary of the TT phase.
The solid lines represent the boundary between the TB and SW phases.
In the upper three lines,  $\eta_{\rm {in}}^*$ is varied at $\eta_{\rm {mb}}^*=0$.
In the bottom three lines,  $\eta_{\rm {mb}}^*$ is varied at $\eta_{\rm {in}}^*=1$.
(b) Stable angle $\theta$ of the TT motion for $\eta_{\rm {in}}^*=1$ and $\eta_{\rm {mb}}^*=0$
as a function of $V^*$ ($\Delta_{\rm S}$ in the inset).
The solid (red) line represent the KS theory \cite{kell82}.
The dashed (blue), dotted (brown), and dashed-dotted (green) lines
represent quasi-spherical perturbation theories at $\dot\gamma\to 0$ by Lebedev {\it et al.} \cite{lebe07}
and  Danker {\it et al.} \cite{dank07}, 
 and at $\dot\gamma\to \infty$ \cite{seif99}, respectively.
The squares and circles indicate the simulation results in Ref. \cite{krau96} and Ref.
{\cite{nogu05}}, respectively.
The dotted (black) lines in the inset represent 
a leading-order approximation of the perturbation theory by Seifert \cite{seif99} (black).
(c) Dependence of $\theta$ and angular velocities on $\eta_{\rm {in}}^*$ at $V^*=0.9$ and  $\eta_{\rm {mb}}^*=0$ obtained by the KS theory.
}
\end{figure}

\section{Vesicle dynamics in steady shear flow} \label{sec:std}

\begin{figure}
\includegraphics{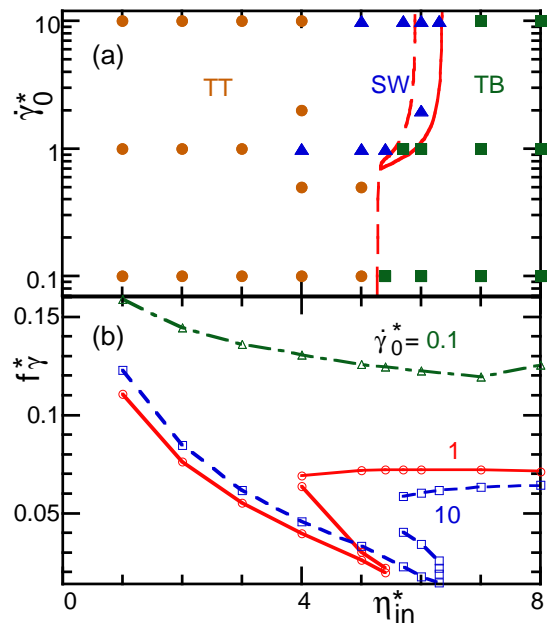}
\caption{ \label{fig:os}
(Color online)
(a) Dynamical phase diagram at $V^*=0.9$, $\eta_{\rm {mb}}^*=0$, and $\dot\gamma_{\rm m}=0$.
The closed circles, squares, and triangles indicate the TT, TB, and SW phases in
the oscillatory flow.
The dashed and solid lines represent the boundary of the TT, TB, and SW phases 
for the steady flow with $\dot\gamma^*=\dot\gamma_0^*$.
(b) Bifurcation lines from one limit cycle to two limit cycles.
The dashed-dotted (green), solid (red), dashed (blue) lines 
represent $\dot\gamma_0=0.1, 1$, and $10$, respectively.
}
\end{figure}

First, we briefly describe the vesicle dynamics in steady shear flow.
Figure \ref{fig:phase}(a) shows 
the phase diagram with dimensionless parameters $S=(7 \pi/3 \sqrt{3}) \dot\gamma \eta_{0}{R_{\rm V}}^3/\kappa \Delta_{\rm S}$
 and $\Lambda= \sqrt{3 \Delta_{\rm S} /10 \pi}(32+ 23\eta_{\rm {in}}^*+ 16\eta_{\rm {mb}}^*)/24$
calculated by the generalized KS theory; see also lines in Fig. \ref{fig:os}(a).
The parameters $S$ and $\Lambda$ are introduced in the perturbation theory of quasi-spherical vesicles \cite{lebe07}.

At low shear rate $\dot\gamma^* \lesssim 1$  ($S \lesssim 1$),
the vesicle keeps a prolate shape, and
 the original KS theory [Eq.(\ref{eq:thetb})] gives very good predictions.
For $B>1$, a stable fixed point $\theta=0.5\arccos(1/B)$ exists,
{\it i.e.}, the TT motion occurs.
For $B<1$, there is no fixed point, and 
the angle $\theta$ periodically rotates (TB).
As $\eta_{\rm {in}}^*$ or $\eta_{\rm {mb}}^*$ ($\Lambda$) increases,
the transition from the TT to TB motion occurs,
where $B$ decreases from $B>1$ to $B<1$.

On the other hand, higher shear at $\dot\gamma^* \gtrsim 1$  ($S \gtrsim 1$)
can induce large shape deformation of the vesicles, and
the SW phase appears between the TT and TB phases.
In Eq.~(\ref{eq:ald}), the shear force depends on $\theta$ as $\sin(2\theta)$,
so the shear increases $\alpha_{\rm D}$ (elongation) for $0<\theta<\pi/2$,
and decreases  $\alpha_{\rm D}$ (shrinkage) for $-\pi/2<\theta<0$.
The SW motion is generated by this shape deformation  as follows \cite{nogu07b}:
A prolate vesicle starts $\theta$ rotation with $B<1$ like in the TB motion,
and then the vesicle shrinks to more spherical shape at $\theta<0$,
which has greater $B$.
When $B$ becomes greater than $1/\cos(2\theta)$,
the right hand side of Eq.(\ref{eq:thetb}) changes its sign and then
the angle $\theta$ increases.
At $\theta>0$, the vesicle elongates to the prolate shape.
Thus, $\theta$ and $\alpha_{\rm D}$ oscillate.

The diagram of these three phases is also obtained in the perturbation theory of quasi-spherical vesicles \cite{lebe07,lebe08,dank07}
and by experiment \cite{desc09}.
All of them give qualitatively the same diagram but quantitative differences exist.
In the perturbation theory by Lebedev {\it et al.} \cite{lebe07,lebe08},
the diagram is determined by only two parameters $S$ and $\Lambda$.
The TT-TB transition occurs at $\Lambda=2/\sqrt{3} \simeq 1.15$ for $S \to 0$.
The SW phase appears between the TT and TB phases at $S>\sqrt{3}$.
The width of the SW phase is small, $\Delta \Lambda_{\rm {SW}} \sim 0.1$
[$1.4 \le \Lambda_{\rm {SW}} \le 1.5$ at $S=10$].
Danker {\it et al.}  \cite{dank07} extended Lebedev's theory to take into account
a higher order term.
In  Danker's theory, the SW phase becomes wider at larger $\Delta_{\rm S}$
($1.4 \le \Lambda_{\rm {SW}}  \le 1.7$ at $S=10$ and $\Delta_{\rm S}=1$).
Very recently, Deschamps {\it et al.} \cite{desc09} obtained the phase diagram experimentally.
Surprisingly, the resulting phase diagram depends only on $\Lambda$ and $S$
for the wide range of the excess area $0.2 \le \Delta_{\rm S} \le 2.2$ ($0.98 \ge V^* \ge 0.79$)
as Lebedev's theory.
However, the width of the SW phase is much wider, $1.5 \le \Lambda_{\rm {SW}} \le 2.2$.
In our generalized KS theory, the SW phase is shifted upwards with decreasing $V^*$
but the region of the SW phase is closer to the experimental results as compared to the perturbation theories;
see Fig. \ref{fig:phase}(a).

To compare the results in more detail,
the stable angle $\theta$ in the TT phase at $\eta_{\rm {in}}^*=1$ and  $\eta_{\rm {mb}}^*=0$ 
is shown in Fig. \ref{fig:phase}(b).
At $V^* \gtrsim 0.9$, several theories \cite{kell82,seif99,dank07,lebe07,nogu07b}, simulations \cite{krau96,nogu05,dupi07}, 
and experiments \cite{abka05,kant05} show good agreements.
Kantsler and Steinberg \cite{kant05} reported that 
their experimental results agree very well with the leading order approximation of the theory \cite{seif99} ($\theta = \pi/4 - C \sqrt{\Delta_{\rm S}} $) with $C \simeq 0.35$ at $V^* \ge 0.82$ ($\Delta_{\rm S} \le 1.8$);
see the black dotted line in the inset of Fig. \ref{fig:phase}(b).
However, at the lower volumes $V^* \lesssim 0.8$ ($\Delta_{\rm S} \gtrsim 2$), 
deviations between the theories are seen.
The KS theory predicts  a gradual decrease in $\theta$, 
while in the perturbation theories, $\theta$ rapidly decreases and 
the TB motion occurs; see Fig. \ref{fig:phase}(b).
The KS theory gives very good agreements with the simulation data
of the boundary integral simulation (without thermal fluctuations) \cite{krau96}
and multi-particle collision (MPC) dynamics (with thermal fluctuations) \cite{nogu05}.
Since the perturbation theory assumes $\Delta_{\rm S} \ll 1$,
this TB phase at  $\eta_{\rm {in}}^*=1$ and $\eta_{\rm {mb}}^*=0$ would be an artifact of the approximation.
It is surprising that the perturbation gives good predictions even at $\Delta_{\rm S} \simeq 1$.
Thus, the perturbation theories should not be applied to the dynamics for  $\Delta_{\rm S} \gtrsim 2$.

Previously, we simulated the SW motion in two- (Ref. \cite{mess09}) and three-dimensional spaces (Ref. \cite{nogu07b})
at a reduced area $A^*=0.7$ and volume $V^*=0.78$  using the MPC method \cite{male99}, respectively (see Fig. \ref{fig:std_snap}).
The region of the SW phase does not completely agree with the prediction of the generalized KS theory
but it can be fitted by modifying the factors $A_0$ or $A_1$.
Since these factors are derived from the perturbation theory,
it is less reliable at low reduced volumes.
In this paper, we use the vesicle at $V^*=0.9$ and $\eta_{\rm {mb}}^*=0$ to investigate the dynamics in the oscillatory flows,
since this condition gives good agreements with experimental results.
The $\eta_{\rm {in}}^*$ dependence of $\theta$ and angular velocities in the KS theory ($\dot\gamma^* \ll 1$) are shown in Fig. \ref{fig:phase}(c).
The mean squared angular velocity $\dot\theta_{\rm m}$ for the rotation is calculated as $\dot\theta_{\rm m}^2 = (1/\pi) \int_0^\pi \dot\theta^2 d\theta$.
At the TT-TB transition point $\eta_{\rm {in}}^* =\eta_{\rm {c}}^* =5.3$,
derivatives of the rotational frequencies or mean angular velocities of the inclination angle $\dot\theta$ and  phase angle $\omega$ are discontinuous,
while $\dot\theta_{\rm m}$ smoothly changes.

\begin{figure}
\includegraphics{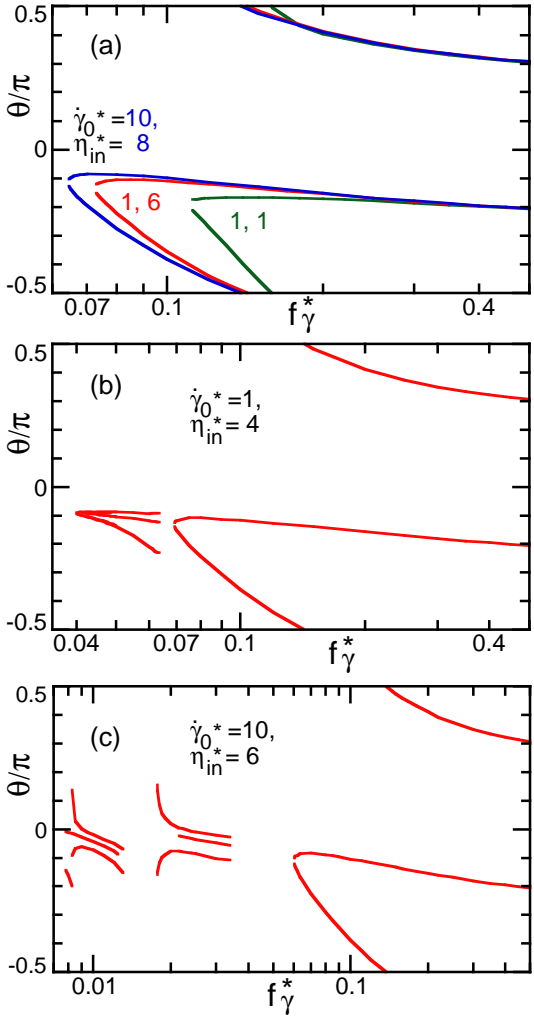}
\caption{ \label{fig:w}
(Color online) 
Domain boundary of the limit-cycle oscillations at (a) 
($\dot\gamma_0^*$, $\eta_{\rm {in}}^*$) = ($1,1$), ($1,6$), and ($10,8$);
(b) ($\dot\gamma_0^*$, $\eta_{\rm {in}}^*$) = ($1,4$); and (c) ($\dot\gamma_0^*$, $\eta_{\rm {in}}^*$) = ($10,6$).
Each domain consists of 
the initial positions $(\alpha_{\rm D},\theta)=(0.325,\theta_{\rm {int}})$ at $t=0$
approaching the same attractor.
}
\end{figure}

\section{Vesicle dynamics in oscillatory shear flow with $\dot\gamma_{\rm m}=0$} \label{sec:osc}

In this section, we describe the vesicle dynamics in the oscillatory shear flow with no net flow ($\dot\gamma_{\rm m}=0$).
Figures \ref{fig:os} and \ref{fig:w} show the phase diagram and domain boundaries of the phases.
The dynamics in the oscillatory flow reflect the dynamics in the steady flow, but
the phase boundary is smeared in the oscillatory flow. 
The SW and TB motions can occur at lower $\eta_{\rm {in}}^*$ than that in the steady flow,
since transient motions in the steady flow can be stabilized in the oscillatory flow.
At low shear frequency $f_\gamma^*$, the vesicle repeats
 the motion for $\eta_{\rm {in}}^*$ in the steady flow with  $\dot\gamma^*  \sim \dot\gamma_0^*$ and $\dot\gamma^*  \sim -\dot\gamma_0^*$.
At a higher frequency $f_\gamma^*$, the vesicle shows two types of oscillations depending on the initial positions in the phase space.
In the middle frequency $f_\gamma^*$, three or more limit-cycle oscillations can coexist at high $\eta_{\rm {in}}^*$ (SW, TB).

\begin{figure}
\includegraphics{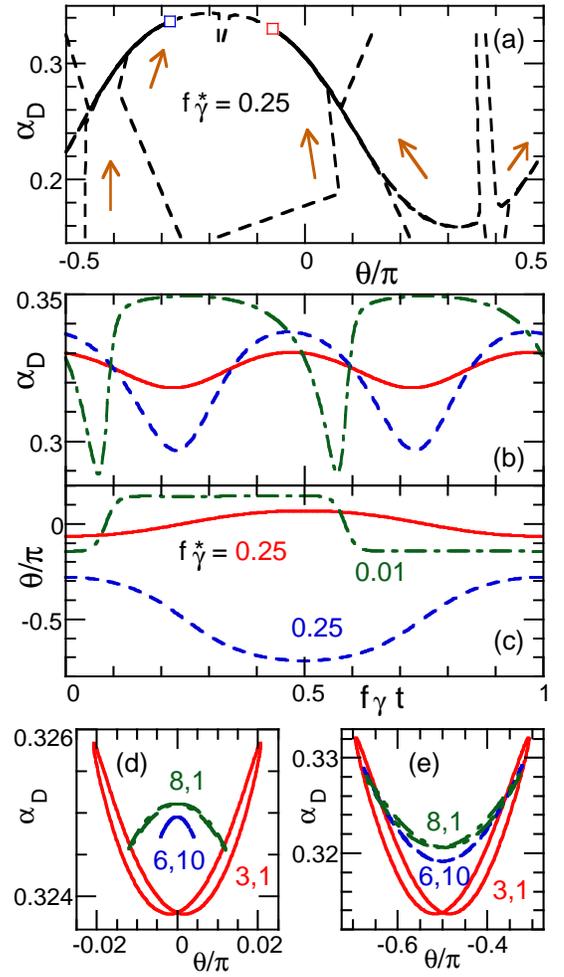}
\caption{ \label{fig:rq_tt}
(Color online)
Vesicle dynamics in the oscillatory flow.
(a)-(c) TT phase at ($\dot\gamma_0^*$, $\eta_{\rm {in}}^*$) = ($1,1$).
(a) Stroboscopic map of the approach to stable fixed points [(blue and red) squares]
from several initial positions on ($\alpha_{\rm D}$, $\theta$) at $t= n/f_\gamma$. The arrows represent the approaching directions.
Time evolutions of $\alpha_{\rm D}$ and $\theta$ in the limit-cycle oscillations
are given in (b) and (c), respectively.
(d), (e) Trajectories of two limit cycles 
at high frequency $f_{\gamma}^*=0.25$
for ($\dot\gamma_0^*$, $\eta_{\rm {in}}^*$)= ($3,1$), ($6,10$), and ($8,1$).
}
\end{figure}

\begin{figure}
\includegraphics{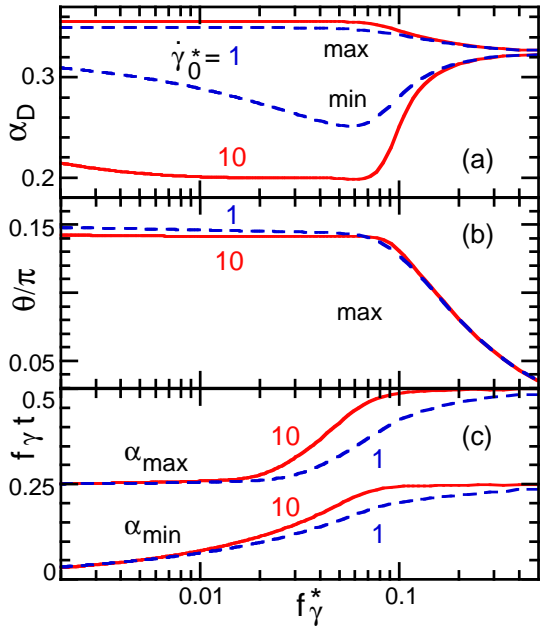}
\caption{ \label{fig:amp_tt}
(Color online)
Dependence on shear frequency $f_\gamma^*$ in the TT phase at $\eta_{\rm {in}}^*=1$:
(a) Maximum and minimum of $\alpha_{\rm D}$,
(b) maximum of $\theta$, and
(c) time $t$ at maximum and minimum of $\alpha_{\rm D}$.
The dashed (blue) and solid (red) lines represent $\dot\gamma^*=1$ and $10$, respectively.
}
\end{figure}

First, we show the dynamics at low $\eta_{\rm {in}}^*$.
For a low shear frequency ($f_{\gamma}^* \lesssim 0.1$),
the vesicle 
 approaches one limit-cycle oscillation from any initial position, and
the angle $\theta$ stepwisely oscillates between $\pm \theta_0$, 
where $\theta_0$ is the stable  angle of the TT phase in the steady flow; see Fig. \ref{fig:rq_tt}(c).
We call this oscillation as TT-based oscillation.
For a high shear frequency ($f_{\gamma}^* \gtrsim 0.1$),
the vesicle cannot relax to the angle $\pm \theta_0$ of the steady flow
for  a half period $1/2f_{\gamma}$.
Instead, two limit cycles coexist as shown in Figs.~\ref{fig:rq_tt}(b) and (c):
Oscillation between $\pm \theta_0$ or between $- \theta_0$ and $\theta_0 -\pi$.
An approached limit cycle is chosen by an initial angle $\theta$
but is almost independent of an initial shape $\alpha_{\rm D}$.
The trajectories on the ($\alpha_{\rm D}$, $\theta$) plane  in the stroboscopic map at $t=n/f_{\gamma}$
are shown in Fig. \ref{fig:rq_tt}(a),
where $n$ is an arbitrary integer.
It rapidly relaxes onto a sinusoidal curve and slowly approaches one of the stable fixed points.
Two saddle points are seen on the curve at $\theta/\pi= -0.18$ and $0.38$.
Figure~\ref{fig:w} shows the domain boundary for initial angles $\theta_{\rm {int}}$.
Since the domain boundary is not sensitive to an initial value of $\alpha_{\rm D}$,
the data with a constant initial shape $\alpha_{\rm D}=0.325$ are shown.
As $f_{\gamma}^*$ decreases, the left stable and unstable fixed points become closer,
and two domains merge at $f_{\gamma}^*=0.11$; see Fig.~\ref{fig:w}(a). 
Thus, the left fixed point (oscillation between $- \theta_0$ and $\theta_0 -\pi$) at high $f_{\gamma}^*$ is 
generated by a saddle-node bifurcation.
The frequency $f_\gamma^*$ for this bifurcation decreases with increasing  $\eta_{\rm {in}}^*$; see Fig. \ref{fig:os}(b).
This can be understood by slower relaxation velocity $\dot\theta_{\rm m}$ at higher $\eta_{\rm {in}}^*$; see Fig. \ref{fig:phase}(c).

\begin{figure}
\includegraphics{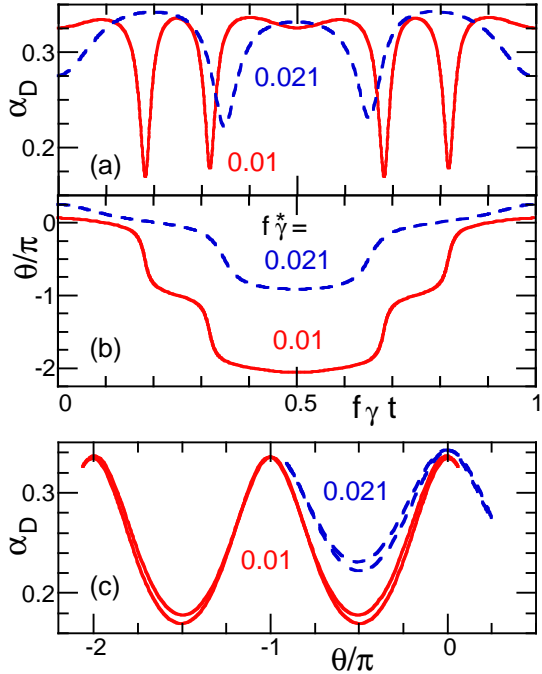}
\caption{ \label{fig:rq_tb}
(Color online)
Time evolutions of $\alpha_{\rm D}$ and $\theta$ in the limit-cycle oscillations 
for the TB phase at ($\dot\gamma_0^*$, $\eta_{\rm {in}}^*$)= ($10,8$).
The solid (red) and dashed (blue) lines represent $f_\gamma^*=0.01$ and $0.021$, respectively.
}
\end{figure}

Figure \ref{fig:amp_tt} shows the dependence of the maxima and minima of $\alpha_{\rm D}$ and $\theta$ in the TT-based oscillation.
Here, we only show the data for $0 \le t \le 0.5/f_\gamma$
because $\alpha_{\rm D}(t+0.5/f_\gamma)=\alpha_{\rm D}(t)$ and $\theta(t+0.5/f_\gamma)=-\theta(t)$.
In the low frequency limit $f_\gamma^* \ll 1$,
the vesicles show maximum and minimum deformation at $t=0.25/f_\gamma$ and $t=0$,
where the shear stress $\eta_0 \dot\gamma(t)$ has maximum and minimum values, respectively.
As $f_\gamma^*$ increases, the times at maximum and minimum deformation
become delayed and approach $t=0.5/f_\gamma$ and  $t=0.25/f_\gamma$.
On the other hand, $\theta$  always has
 maximum and minimum values at $t=0.5/f_\gamma$ and $t=0$, respectively.
The amplitudes of  $\alpha_{\rm D}$ and $\theta$ in the oscillation 
rapidly decrease with increasing $f_\gamma^*$ at  $f_\gamma^* \gtrsim 0.1$.
This reduction of amplitudes and the coexistence of two limit cycles at
high $f_\gamma^*$ occur due to faster change of shear direction than vesicle relaxation.

\begin{figure}
\includegraphics{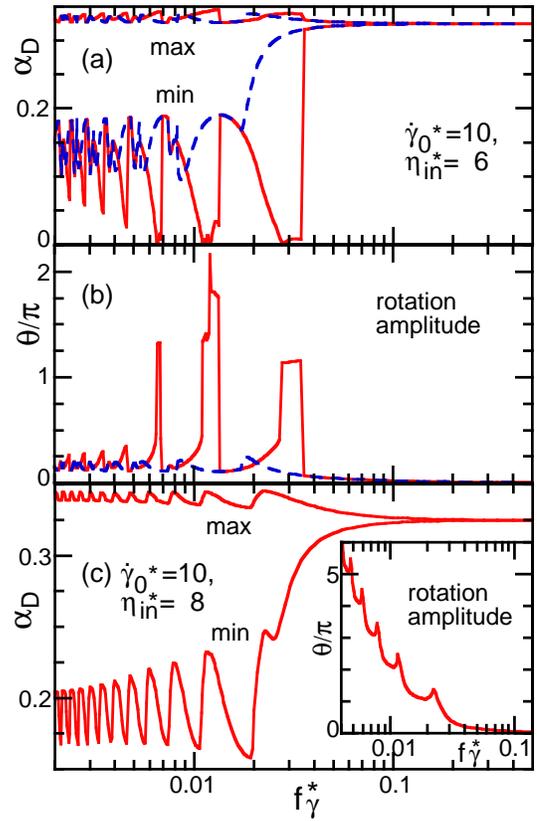}
\caption{ \label{fig:amp_sw}
(Color online)
Dependence on shear frequency $f_\gamma^*$ in the (a), (b) SW and (c) TB phases at 
 ($\dot\gamma_0^*$, $\eta_{\rm {in}}^*$)= ($10,6$) and ($10,8$), respectively.
(a), (c)  Maximum and minimum of $\alpha_{\rm D}$.
The peak-to-peak amplitudes of $\theta$ in the oscillations are
shown in (b) and the inset of (c).
The solid (red) and dashed (blue) lines in (a) and (b) represent the data 
obtained with increasing or decreasing $f_\gamma^*$, respectively.
}
\end{figure}

At low $\eta_{\rm {in}}^*$ [(brown) circles in  Fig. \ref{fig:os}(a)],
one region of two limit cycles appears only at high $f_\gamma^*$.
On the other hand, at higher $\eta_{\rm {in}}^*$ [(blue) triangles and (green) squares in  Fig. \ref{fig:os}(a)],
regions of two or more limit cycles appear at lower $f_\gamma^*$.
Next, we describe TB-based oscillation at a high $\eta_{\rm {in}}^*$  [(green) squares in  Fig. \ref{fig:os}(a)].
At low $f_\gamma^*$, the vesicle rotates clockwisely at $0 < t < 0.5/f_\gamma$
and rotates back at $0.5/f_\gamma < t < 1/f_\gamma$; see  Fig. \ref{fig:rq_tb}.
Typically, the shape of time-evolution curves are symmetric like the TT-based oscillation,
but asymmetric shape (see dashed lines in Fig. \ref{fig:rq_tb}) and
the coexistence of two limit cycles can occur even at low $f_\gamma^*$.
The rotation amplitude of $\theta$ stepwisely decreases with increasing $f_\gamma^*$
and the amplitude of $\alpha_{\rm D}$ oscillates  as shown in  Fig. \ref{fig:amp_sw}(c).
At peaks of rotation amplitude in the inset of Fig. \ref{fig:amp_sw}(c),
two limit cycles with different rotation amplitudes 
are obtained with increasing and decreasing $f_\gamma^*$.
Thus, narrow regions of two limit cycles repeatedly appear at $f_\gamma^* \lesssim 0.1$.
At high $f_\gamma^*$, a wide region  of  two limit cycles appear like the TT-based oscillation.
The trajectories of the oscillation around $\theta/\pi=-0.5$ are also very similar; see  Fig. \ref{fig:rq_tt}(e).
However, the trajectories around $\theta/\pi=0$ are convex upward and downward
for the TB- and TT-based oscillations, respectively, (see  Fig. \ref{fig:rq_tt}(d)),
since $\theta$ rotates in the opposite direction.

\begin{figure}
\includegraphics{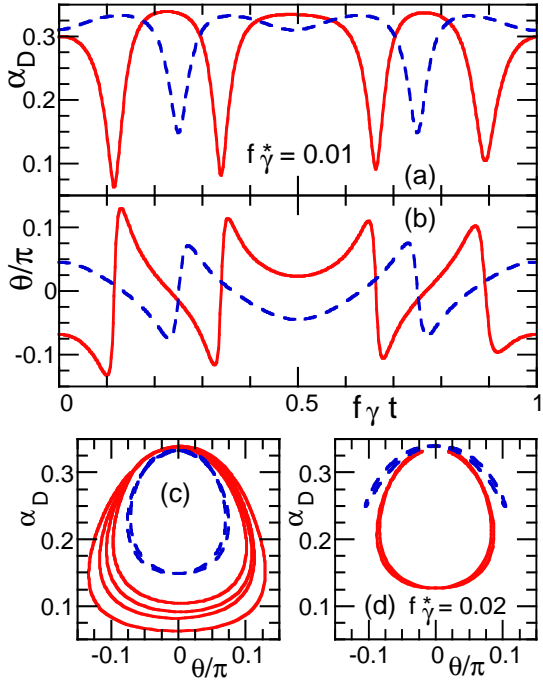}
\caption{ \label{fig:rq_sw}
(Color online)
Time evolutions of $\alpha_{\rm D}$ and $\theta$ in the limit-cycle oscillations 
for the SW phase at ($\dot\gamma_0^*$, $\eta_{\rm {in}}^*$)= ($10,6$)
at (a)-(c) $f_\gamma^*=0.01$ and (d) $0.02$.
Two coexisting limit cycles are shown as the solid (red) and dashed (blue) lines.
}
\end{figure}

At middle $\eta_{\rm {in}}^*$ and high $\dot\gamma^*$ [(blue) triangles in  Fig. \ref{fig:os}(a)],
the SW-based oscillation appears.
The different frequencies of oscillations of $\alpha_{\rm D}$ and $\theta$
occur in a wider region of $f_\gamma^*$ than the TB region; see  Fig. \ref{fig:rq_sw}.
The domains of multiple limit cycles have a complicated shape, as shown in Fig. \ref{fig:w}(c).
These domains appear or disappear via saddle-node and Hopf bifurcations.
The amplitudes of  $\alpha_{\rm D}$ and $\theta$ depend on trapped limit cycles.
As $f_\gamma^*$ increases or decreases, the vesicles are often trapped in different limit cycles;
see Figs. \ref{fig:amp_sw}(a) and (b).
The peaks in Fig. \ref{fig:amp_sw}(b) represent TB rotation, which coexists with SW oscillation.

\begin{figure}
\includegraphics{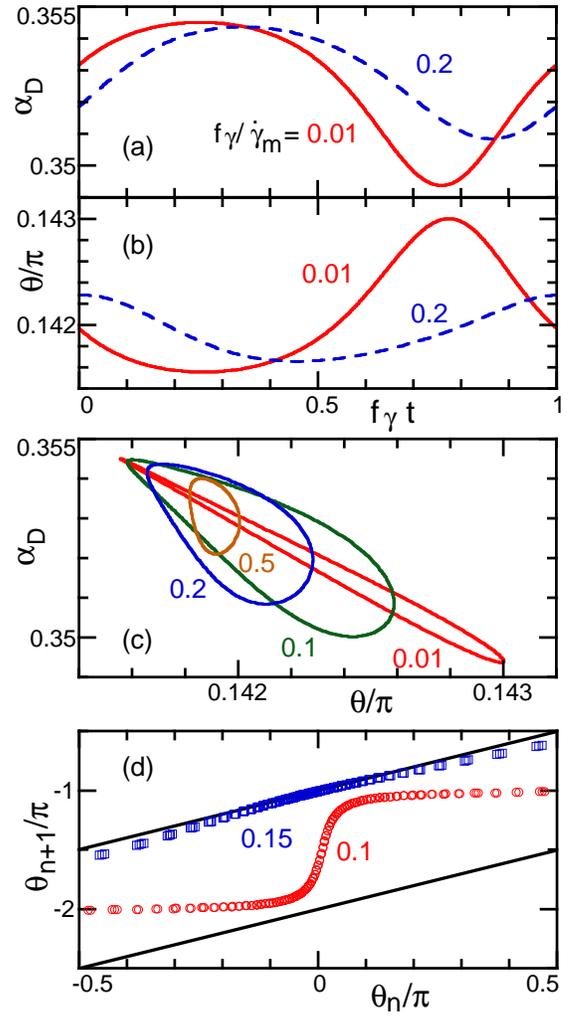}
\caption{ \label{fig:bias_rq}
(Color online)
Vesicle dynamics for the TT phase in the oscillatory flow with finite mean shear rates $\dot\gamma_{\rm m}$.
(a-c) TT phase at $\eta_{\rm {in}}^*=1$, $\dot\gamma_{\rm m}^*=2$, and
 $\dot\gamma_0/\dot\gamma_{\rm m}=0.5$.
(a), (b) Time evolutions of $\alpha_{\rm D}$ and $\theta$ in the limit-cycle oscillations.
The solid (red) and dashed (blue) lines represent $f_\gamma^*=0.01$ and $0.2$, respectively.
(c) Trajectories at  $f_\gamma/\dot\gamma_{\rm m}=0.01$, $0.1$, $0.2$, and $0.5$.
(d) Return map sampled stroboscopically at $t=n/f_\gamma$  in the TB attractor at $\eta_{\rm {in}}^*=8$, $\dot\gamma_{\rm m}^*=10$, and
 $\dot\gamma_0/\dot\gamma_{\rm m}=5$.
Circles (red) and squares (blue) represent $f_\gamma/\dot\gamma_{\rm m}=0.1$ and $0.15$, respectively.
Solid (black) lines show $\theta_{n+1}=\theta_n -\pi$ and  $\theta_{n+1}=\theta_n -2\pi$.
}
\end{figure}

\begin{figure}
\includegraphics{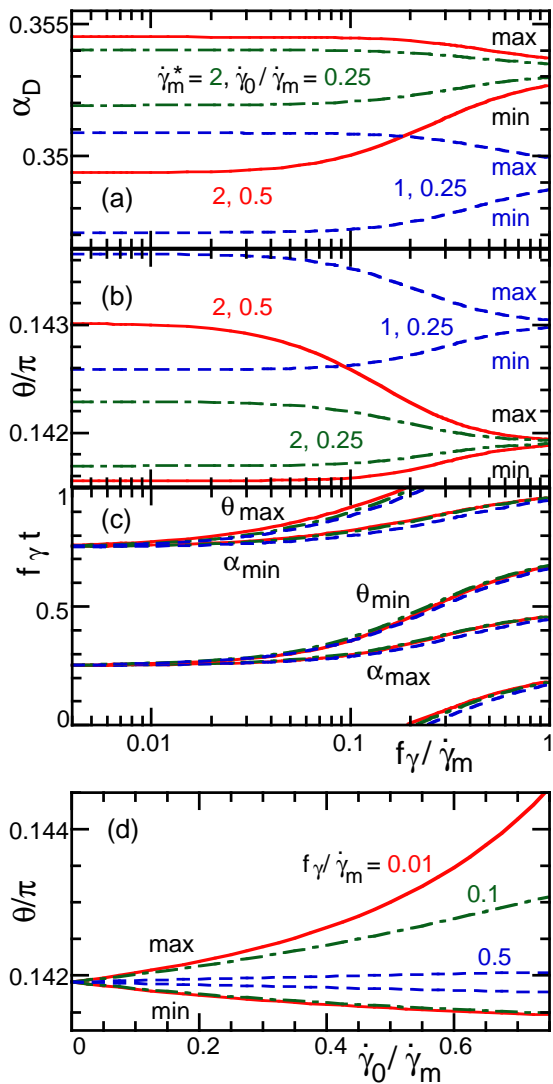}
\caption{ \label{fig:bias_amp}
(Color online)
(a)-(c) Dependence on the frequency $f_\gamma^*$ for the TT phase in the oscillatory flow with
$\dot\gamma_{\rm m}^*=1$ or $2$:
Maximum and minimum of (a) $\alpha_{\rm D}$ and (b) $\theta$; and
(c) time $t$ at maximum and minimum of $\alpha_{\rm D}$ and $\theta$.
The solid (red), dashed-dotted (green), and dashed (blue) lines represent 
$(\dot\gamma_{\rm m}^*, \dot\gamma_0/\dot\gamma_{\rm m})=  (2,0.05)$, $(2,0.25)$, and $(1,0.25)$, respectively.
(d) Shear-amplitude $\dot\gamma_0$ dependence of the maximum and minimum of $\theta$
at $\dot\gamma_{\rm m}^*=2$.
The solid (red), dashed-dotted (green), and dashed (blue) lines represent 
$f_\gamma^*=0.01$, $0.1$, and $0.5$, respectively.
}
\end{figure}

\section{Vesicle dynamics in oscillatory shear flow with finite $\dot\gamma_{\rm m}$} \label{sec:bias}

Next, we consider the vesicle dynamics in the oscillatory flow with  $0< \dot\gamma_0 < \dot\gamma_{\rm m}$.
In the TT phase,
the shear rate $\dot\gamma$ is always positive and the vesicle keeps a stable TT angle with $\theta>0$.
Unlike the oscillatory flow with $\dot\gamma_{\rm m}=0$, only one limit cycle exists even for high $f_\gamma$.
The shape $\alpha_{\rm D}$ and angle $\theta$ oscillate around the steady values at $\dot\gamma(t)=\dot\gamma_{\rm m}$; 
see Fig. \ref{fig:bias_rq}. 
Their amplitudes increase with increasing $\dot\gamma_0$ and decrease with increasing $f_\gamma$; see Fig. \ref{fig:bias_amp}.
In the low frequency limit,
the maximum and minimum of $\alpha_{\rm D}$ (minimum and maximum of $\theta$) appear at $t=0.25/f_\gamma$ and $t=0.75/f_\gamma$,
where the shear stress $\eta_0 \dot\gamma(t)$ has maximum and minimum values, respectively.
The minimum and maximum times are delayed with increasing $f_\gamma$, and
these delays are almost independent of $\dot\gamma_0$ and $\dot\gamma_{\rm m}$; see Fig. \ref{fig:bias_amp}(c).
In the high frequency limit, the maximum and minimum of $\alpha_{\rm D}$ are reached  at $t=0.5/f_\gamma$ and $t=1/f_\gamma$, respectively,
where $\dot\gamma=\dot\gamma_{\rm m}$.
The angle $\theta$ shows greater delays than  $\alpha_{\rm D}$,
since stable $\theta$ is varied not directly by $\dot\gamma^*$ but by the evolution of $\alpha_{\rm D}$.
The phase delay between  $\alpha_{\rm D}$ and $\theta$ approaches $\pi/4$ as $f_\gamma/\dot\gamma_{\rm m}$ increases; see  Fig. \ref{fig:bias_rq}(c). 
These oscillations also occur for RBCs, where the TT motion generates additional oscillation of the shape and angles \cite{nogu09c}.
A similar phase delay is observed in the experiments of RBCs for oscillatory shear flow \cite{naka90}.

In the TB and SW phases, the shear oscillation is weakly coupled with the oscillations of $\alpha_{\rm D}$ and $\theta$.
In most of the parameter regions, the TB or SW oscillation is not synchronized with the shear oscillation.
The synchronization only occurs in very narrow regions, where the tumbling or swinging frequency has
integer ratios with the shear frequency: $f_{\rm {TB}} = (n/m) f_\gamma$ or $f_{\rm {SW}} = (n/m) f_\gamma$,
where $m$ and $n$  are arbitrary integers \cite{berg84}.
The synchronization with $f_{\rm {TB}} = f_\gamma$ and $f_{\rm {TB}} = 2f_\gamma$
 appears only at $0.1522<f_\gamma/\dot\gamma_{\rm m}<0.1542$ and at $0.0764<f_\gamma/\dot\gamma_{\rm m}<0.0768$, respectively,
for $\eta_{\rm {in}}^*=8$, $\dot\gamma_{\rm m}^*=10$, and $\dot\gamma_0^*=5$.
Other synchronizations have much narrower regions.
Figure \ref{fig:bias_rq}(d) shows the return map of typical attracted orbits in the TB phases.
At $\dot\gamma_0/\dot\gamma_{\rm m}=0.15$, 
an intermittent dynamics appears;
$\theta_{n+1}$ ($\theta$ at $t=(n+1)/f_\gamma$) approaches $\theta_{n+1}=\theta_n-\pi$ at $\theta_n=0$ (but not contacted),
so that the vesicle is trapped on $\theta=0$ at $t=n/f_\gamma$
and  intermittently escapes.

Elastic capsules such as RBCs \cite{abka07} and synthetic capsules \cite{chan93,walt01,kess08,sui08,bagc09}
show similar oscillations of $\alpha_{\rm D}$ and $\theta$ at the tank-treading phase in steady shear flow.
To overcome the energy barrier,
the TT membrane velocity $\omega$ of the elastic capsules oscillates with the TT rotation frequency,
 and induces the oscillations of $\alpha_{\rm D}$ and $\theta$ \cite{skot07,nogu09b}.
For the elastic capsules,
the phase difference between  $\alpha_{\rm D}$ and $\theta$ is also $\pi/4$ at a large shear rate,
but approaches $0$ at lower bound $\dot\gamma_{\rm {tt}}$ of the TT phase \cite{nogu09b}.
This phase shift occurs since
$\theta$ oscillation is caused by $\omega$ oscillation as well as $\alpha_{\rm D}$ oscillation
unlike for the lipid vesicles in the oscillatory flow.

\section{Effects of thermal fluctuations in oscillatory shear flow} \label{sec:jef}

\begin{figure}
\includegraphics{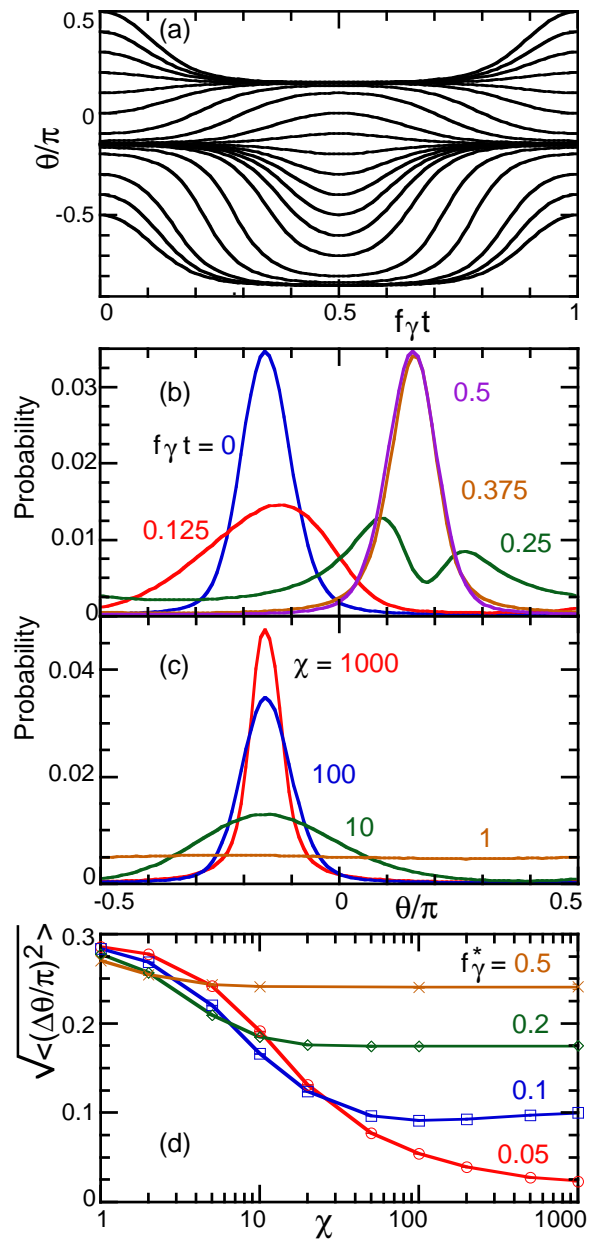}
\caption{ \label{fig:ttwn}
(Color online)
Dynamics of vesicles with a fixed prolate shape at $V^*=0.9$ and $\eta_{\rm {in}}^*=1$ 
in the oscillatory flow with $\dot\gamma_{\rm m}=0$.
(a) Time evolution of $\theta$ without thermal fluctuations for $f_{\gamma}^*=0.1$.
(b) Time evolution of probability distribution of $\theta$ at
 Peclet number $\chi=100$ and $f_{\gamma}^*=0.1$.
(c) Probability distribution of $\theta$ at $t=n/f_\gamma$ for 
 various $\chi$ with  $f_{\gamma}^*=0.1$.
(d) Peclet number $\chi$ dependence of the standard deviation
of $\theta$ distribution  at $t=n/f_{\gamma}$
 for various $f_{\gamma}^*$.
The error bars are smaller than symbols.
}
\end{figure}

In the small shear limit $\dot\gamma^* \to 0$,
the vesicle deformation becomes negligible and
the vesicle dynamics is given by the original KS theory,
{\it i.e.}, $\alpha_{\rm D}$ and the factor $B$ in
Eq. (\ref{eq:thetb}) are constant.
The  original KS theory gives no attractor in the oscillatory flow with $\dot\gamma_{\rm m}^*=0$.
Since Eq. (\ref{eq:thetb}) is symmetric to time reversal,
$\theta$  exactly returns to the original angle at $t'=t+n/f_\gamma$; see Fig. \ref{fig:ttwn}(a).
The shape deformation given by Eq. (\ref{eq:ald}) breaks this reversibility
but the approach to the limit cycle is slower at smaller $\dot\gamma_0^*$. 
At small $\dot\gamma_0^*$, the thermal noises are typically not negligible.
Recently, Pine {\it et al.} \cite{pine05} demonstrated that noises can break the time reversibility:
The interactions between spherical particles can induce the diffusion of particle positions at $t=n/f_\gamma$
in the oscillatory shear flow.
In this section, we clarify the thermal noise effects on the vesicle dynamics with a fixed shape.

Gaussian white noise $g(t)$ is added to Eq. (\ref{eq:thetb})
to take into account the thermal fluctuations, where
$\langle g(t)\rangle = 0$,
$\langle g(t)g(t')\rangle = 2 D \delta(t-t')$.
The fluctuation-dissipation theorem gives
the diffusion constant $D=k_{\rm B}T/\zeta$,
where $\zeta$ is a rotational friction coefficient and $k_{\rm B}T$ is the thermal energy;
for a sphere, $\zeta=8\pi \eta_{\rm {0}}R_{\rm S}^3$.
A dimensionless quantity, the rotational Peclet number
$\chi = \dot\gamma_0/D=\dot\gamma_0\zeta/k_{\rm B}T$
represents the shear amplitude relative to the thermal fluctuations.
Equation~(\ref{eq:thetb})
is numerically integrated
using the second-order Runge-Kutta method~\cite{hone92} with a time step $\Delta t \leq 0.00005$.

\begin{figure}
\includegraphics{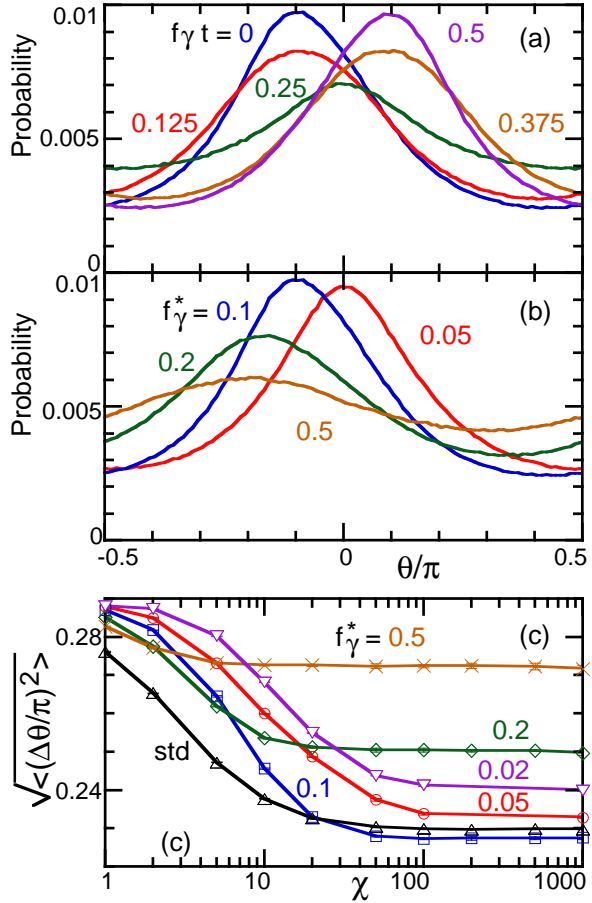}
\caption{ \label{fig:jefwn}
(Color online)
Dynamics of a solid prolate object ($\eta_{\rm {in}}^*\to \infty$) at $V^*=0.9$ in the oscillatory flow with $\dot\gamma_{\rm m}=0$.
(a) Time evolution of probability distribution of $\theta$ at
 Peclet number $\chi=1000$ and $f_{\gamma}^*=0.1$.
(b) Probability distribution of $\theta$ at $t=n/f_\gamma$, $\chi=1000$,
 and  $f_{\gamma}^*=0.05, 0.1, 0.2, 0.5$.
(c) Peclet number $\chi$ dependence of the standard deviation
of $\theta$ distribution  at $t=n/f_{\gamma}$ for various $f_{\gamma}^*$.
The data ($\triangle$) for steady flow with $\chi=\dot\gamma/D$ are also shown.
The error bars are smaller than symbols.
}
\end{figure}

First, we consider the TT vesicle at $\eta_{\rm {in}}^* =1$ 
($B=1.7$, where the stable TT angle $\theta_0=0.15\pi$ in the steady flow).
At $\chi \lesssim 1$, the thermal fluctuations are dominant, 
and $\theta$ show a uniform distribution.
As $\chi$ increases,  a peak grows at $\theta \simeq -\theta_0$ or at $\theta \simeq \theta_0$
in the distribution of $\theta$ at $t=n/f_\gamma$ or at $t=(n+0.5)/f_\gamma$, respectively; see Figs. \ref{fig:ttwn}(b) and (c).
For $n/f_\gamma <t< (n+0.5)/f_\gamma$,
 a half of the vesicles rotate
 $\theta \simeq -\theta_0$ to $\theta_0$ and the other half rotate to $\theta_0-\pi$.
Thus, the peak at $\theta \simeq -\theta_0$ splits to two peaks,
and then they merge into a new peak at $\theta \simeq \theta_0$.
Thus, the thermal fluctuations lead the vesicle to the orbit
of the deformed vesicle; compare Fig. \ref{fig:rq_tt}(c).
These effects can be explained by the orbits in Fig. \ref{fig:ttwn}(a).
The angles in the wide region of  $-0.8<\theta/\pi <0.15$ at  $t=(n+0.5)/f_\gamma$
accumulate in the narrow region $-0.2<\theta/\pi <0.1$ at $t=(n+1)/f_\gamma$.
The combination of this accumulation and the thermal diffusion
gives the peaks in the angular distribution.
The width of the distribution is calculated as
the standard deviation $\sqrt{\langle(\Delta\theta/\pi)^2\rangle}$ 
for $\theta_{\rm {max}}-0.5\pi<\theta<\theta_{\rm {max}}+0.5\pi$,
where $\theta_{\rm {max}}$ is the maximum angle; see  Fig. \ref{fig:ttwn}(d).
For the uniform distribution, $\sqrt{\langle(\Delta\theta/\pi)^2\rangle}=1/2\sqrt{3} \simeq 0.29$.
At $\chi \gg 1$, the width decreases with decreasing $f_\gamma^*$.
A sharper peak is obtained for smaller $f_\gamma^*$ with large $\chi$.

Next, we investigate the vesicle in the TB phase.
In the limit $\eta_{\rm {in}}^* \to \infty$, 
the vesicle becomes a solid object,
where no shape deformation occur even at high $\dot\gamma^*$.
Then, the dynamics is given by Jeffery's theory \cite{jeff22},
where $B=f_0 f_1$ in  Eq. (\ref{eq:thetb})
($B=0.59$ for a prolate object at $V^*=0.9$).
In the absence of the thermal noise, 
the orbit is time reversible and
the angular distribution has a peak at $\theta=0$ in the steady flow.
The thermal fluctuations slightly shift this peak to positive $\theta$ \cite{nogu05},
and the peak becomes smaller at smaller $\chi$; see Fig. \ref{fig:jefwn}(c).
In the oscillatory flow, 
the peak in the angular distribution moves to negative (positive) $\theta$
for $t=n/f_\gamma$ ($t=(n+0.5)/f_\gamma$) with increasing $f_\gamma^*$; see Fig. \ref{fig:jefwn}.
For $n/f_\gamma <t< (n+0.5)/f_\gamma$,
 most of the TB vesicles show 
clockwise rotation from $\theta \sim -\pi/4$ to  $-3 \pi/4$,
while the peak moves from $-\pi/4$ to $\pi/4$.
The sharpest peak is obtained for $f_\gamma^* \simeq 0.1$ with large $\chi$ 
in contract to a monotonic dependence on $f_\gamma^*$ in TT phase; 
compare Figs. \ref{fig:ttwn}(d) and \ref{fig:jefwn}(c).
At smaller frequency $f_\gamma^* \ll 0.1$, 
it approach the uniform distribution.
Thus, very small thermal noise can break the time reversibility for both the TT and TB phases
and generate
a preferred orbit, although the relaxation time is longer for the smaller noise.

\section{Summary}~\label{sec:sum}

We found the fruitful dynamics of fluid vesicles in oscillatory shear flow.
For no net flow ($\dot\gamma_{\rm m}=0$),
the vesicle approaches one or multiple limit-cycle oscillations.
In the TT phase (low inertial viscosity $\eta_{\rm {in}}^*$)
at  a low shear frequency $f_\gamma^*$,
the inclination angle $\theta$ oscillates between the steady TT angle $\pm \theta_0$ 
in the steady flow with $\dot\gamma^* \sim \pm \dot\gamma_0^*$
at $n<f_\gamma t<n+0.5$ and $n+0.5<f_\gamma t<n+1$, respectively.
In the TB phase (high $\eta_{\rm {in}}^*$),
$\theta$ rotates clockwisely and then rotates back.
In the SW phase  (middle $\eta_{\rm {in}}^*$ and high $\dot\gamma_0^*$),
the vesicle shape and $\theta$ oscillate.
At high $f_\gamma^*$, two limit cycles coexist.
The angle $\theta$ oscillates around $\theta =0$ or $\theta=-\pi/2$.
At some regions of middle $f_\gamma^*$, two or more limit cycles can exist
in the TB and SW phases.

In the oscillatory flow with the finite net flow at $\dot\gamma_{\rm m}>\dot\gamma_0>0$,
only one limit cycle exists in the TT phase.
As $f_\gamma^*$ increases, the oscillation of shape and $\theta$ become delayed 
compared to the shear oscillation.
The SW and TB oscillations are typically not locked with the shear frequency.
In the limit of small $\dot\gamma_0^*$ or high $\eta_{\rm {in}}^*$ (corresponding to solid objects),
the dynamics is integrable and no attractor exists in the absence of noise.
However, we found that the thermal fluctuations can induce an attractive orbit. 

In the previous paper \cite{nogu09b}, we investigated the dynamics of an RBC in oscillatory shear flow
using the differential equations of three variables.
RBCs show TT, TB, and intermediate motions depending on $\dot\gamma_0^*$ and $\eta_{\rm {in}}^*$.
In the TT and TB phases, the dynamics of RBCs is very similar to that of fluid vesicles,
while the membrane energy barrier of RBCs induces additional shape and angular oscillations.
For RBCs with the intermediate shear amplitude,  the small variations in the parameters can induce the dynamic modes
from intermittent to synchronized oscillations, and change the number of coexisting limit-cycle oscillations.

Recently, Deschamps {\it et al.} \cite{desc09a} demonstrated that a general flow controlled by a 4-roll mill device
is a good experimental setup to investigate the details of the vesicle dynamics.
Our study shows that the oscillatory shear flow is another condition to reveal the dynamics of vesicles, cells, and capsules.
The analysis of the frequency dependence helps understand their viscoelastic behaviors.
We hope that our study stimulates further experimental study on oscillatory flows.

\section*{Acknowledgment}
We  would like to thank 
G. Gompper (J{\"u}lich) for the helpful discussion.
This study is partially supported by a Grant-in-Aid for Scientific Research on Priority Area ``Soft Matter Physics'' from
the Ministry of Education, Culture, Sports, Science, and Technology of Japan.


\begin{thebibliography}{60}

\bibitem{rall84}
J. M. Rallison: Ann.\ Rev.\ Fluid\ Mech. \textbf{16} (1984) 45. 

\bibitem{ston94}
H.~A. Stone: Ann.\ Rev.\ Fluid\ Mech. \textbf{26} (1994) 65.

\bibitem{haas97}
K.~H. de~Haas, C. Blom, D. van~den Ende, M.~H.~G. Duits, and J. Mellema: Phys.\ Rev. E \textbf{56} (1997) 7132.

\bibitem{abka05}
M. Abkarian and A. Viallat: Biophys.\ J. \textbf{89} (2005) 1055.

\bibitem{made06}
M.~A. Mader, V. Vitkova, M. Abkarian, A. Viallat and T. Podgorski: Eur.\ Phys. J.\ E \textbf{19} 
   ({2006}) {389}.

\bibitem{kant05}
V. Kantsler {and} {V.} {Steinberg}: Phys.\ Rev.\ Lett.  \textbf{{95}} ({2005}) {258101}.

\bibitem{kant06}
V. Kantsler {and} {V.} {Steinberg}: Phys.\ Rev.\ Lett.  \textbf{{96}} ({2006}) {036001}.

\bibitem{desc09}
J. Deschamps, V. Kantsler, {and} {V.} {Steinberg}: Phys.\ Rev.\ Lett.  \textbf{{102}} ({2009}) {118105}.

\bibitem{desc09a}
J. Deschamps, V. Kantsler, E. Segre, {and} {V.} {Steinberg}: Proc.\ Natl.\ Acad.\ Sci.\ USA \textbf{{106}} ({2009}) {11444}.

\bibitem{kant07}
V. Kantsler, E. Segre, {and} {V.} {Steinberg}: Phys.\ Rev.\ Lett.  \textbf{{99}} ({2007}) {178102}.

\bibitem{krau96} {M.} {Kraus}, {W.} {Wintz}, {U.} {Seifert}, {and} {R.} {Lipowsky}: {Phys.\ Rev.\ Lett.} \textbf{{77}} ({1996}) {3685}.

\bibitem{seif99} {U.} {Seifert}: {Eur.\ Phys.\ J. B} \textbf{{8}} ({1999}) {405}.

\bibitem{misb06} {C.} {Misbah}: {Phys.\ Rev.\ Lett.} \textbf{{96}} ({2006}) {028104}.

\bibitem{dank07}
G. Danker, T. Biben, T. Podgorski, C. {Verdier}, and
C. Misbah: Phys. Rev. E  \textbf{{76}} ({2007}) {041905}.

\bibitem{lebe07} {V.~V.} {Lebedev}, {K.~S.} {Turitsyn}, {and} {S.~S.} {Vergeles}: {Phys.\ Rev.\ Lett.} \textbf{{99}} ({2007}) {218101}.

\bibitem{lebe08} {V.~V.} {Lebedev}, {K.~S.} {Turitsyn}, {and} {S.~S.} {Vergeles}: New.\ J.\ Phys. \textbf{{10}} ({2008}) {043044}.

\bibitem{turi08} {K.~S.} {Turitsyn}, {and} {S.~S.} {Vergeles}: {Phys.\ Rev.\ Lett.} \textbf{{100}} ({2008}) {028103}.

\bibitem{vlah07}
P.~M. Vlahovska {and} {R.~S.} {Gracia}: Phys.\ Rev.\ E \textbf{{75}} ({2007}) {016313}.

\bibitem{nogu04} {H.} {Noguchi} {and} {G.} {Gompper}: {Phys.\ Rev.\ Lett.} \textbf{{93}} ({2004}) {258102}.

\bibitem{nogu05} {H.} {Noguchi} {and} {G.} {Gompper}: {Phys.\ Rev.\ E} \textbf{{72}} ({2005}) {011901}.

\bibitem{nogu07b} {H.} {Noguchi} {and} {G.} {Gompper}: {Phys.\ Rev.\ Lett.} \textbf{{98}} ({2007}) {128103}.

\bibitem{nogu09} {H.} {Noguchi}: J.\ Phys.\ Soc.\ Jpn. \textbf{{78}} ({2009}) {041007}.

\bibitem{mess09}  {S.} {Me{\ss}linger}, {B.} {Schmidt}, {H.} {Noguchi}, {and} {G.} {Gompper}:  {Phys.\ Rev.\ E} 
  \textbf{{80}} ({2009}) {011901}.

\bibitem{skal90} {R.} {Skalak}: {Biorheology} \textbf{{27}} ({1990}) {277}.

\bibitem{fung04} {Y.~C.} {Fung}:
  \emph{{Biomechanics: mechanical properties of living tissues}}
  ({Springer}, {Berlin}, {2004}), {2nd} ed.

\bibitem{fisc78} {T.~M.} {Fischer}, {M.} {St{\"o}hr-Liesen}, {and}
{H.} {Schmid-Sch{\"o}nbein}: Science  \textbf{202} (1978) 894.

\bibitem{abka07} {M.} {Abkarian}, {M.} {Faivre}, {and} {A.} {Viallat}: {Phys.\ Rev.\ Lett.} \textbf{{98}} ({2007}) {188302}.

\bibitem{abka08a} {M.} {Abkarian}, {M.} {Faivre}, {R.} {Horton}, {K.} {Smistrup}, {C.~A.} {Best-Popescu}, {and} {H.~A.} {Stone}:
Biomed. Mater.  \textbf{{3}} ({2008}) {034011}.

\bibitem{kell82} {S.~R.} {Keller} {and} {R.} {Skalak}: {J.\ Fluid\ Mech.} \textbf{{120}} ({1982}) {27}.

\bibitem{tran84} {R.} {Tran-Son-Tay}, {S.~P.} {Sutera}, and {P.~R.} {Rao}: {Biophys.\ J.} \textbf{{46}} ({1984}) {65}.

\bibitem{naka90} {T.} {Nakajima}, {K.} {Kon}, {N.} {Maeda}, {K.} {Tsunekawa}, {and} {T.} {Shiga}:
Am.\ J.\ Physiol. \textbf{{259}} ({1990}) {H1071}.

\bibitem{wata06}
{N.} {Watanabe}, {H.} {Kataoka}, {T.} {Yasuda}, {and} {S.} {Takatani}: Biophys.\ J. \textbf{{91}} ({2006}) {1984}.

\bibitem{pozr03} {C.} {Pozrikidis}: Annals\ Biomed.\ Eng. \textbf{{31}} ({2003}) {1194}.

\bibitem{pozr05} {C.} {Pozrikidis}: Phys. Fluids \textbf{{17}} ({2005}) {031503}.

\bibitem{skot07} {J.~M.} {Skotheim} {and} {T.~W.} {Secomb}: {Phys.\ Rev.\ Lett.} \textbf{{98}} ({2007}) {078301}.

\bibitem{dupi07} {M.~M.} {Dupin}, {I.} {Halliday}, {C.~M.} {Care}, {L.} {Alboul}, {and} {L.~L.} {Munn}: {Phys.\ Rev.\ E} \textbf{{75}} ({2007}) {066707}.

\bibitem{macm09} {R.~M.} {MacMeccan}, {J.~R.} {Clausen}, {G.~P.} {Neitzel}, {and} {C.~K.} {Aidun}: J. Fluid. Mech.
  \textbf{{618}} ({2009}) {13}.

\bibitem{nogu05b} {H.} {Noguchi} {and} {G.} {Gompper}: {Proc.\ Natl.\ Acad.\ Sci.\ USA} \textbf{{102}} ({2005}) {14159}.

\bibitem{mcwh09} {J. L.} {McWhirter}, {H.} {Noguchi}, {and} {G.} {Gompper}: Proc.\ Natl.\ Acad.\ Sci.\ USA \textbf{106} (2009) {6039}.

\bibitem{nogu09b} {H.} {Noguchi}: Phys. Rev. E  \textbf{{80}} ({2009}) {021902}.

\bibitem{nogu09c} {H.} {Noguchi}: {arXiv:0903.0038 [cond-mat.soft]}.

\bibitem{chan93} {K. S.} {Chang} {and} {W.~L.} {Olbricht}: {J. Fluid Mech.} \textbf{{250}} ({1993}) {609}.

\bibitem{walt01} {A.} {Walter}, {H.} {Rehage}, {and} {H.} {Leonhard}: Colloids Surf. A \textbf{183-185} (2001) 123.

\bibitem{kess08} {S.} {Kessler} {R.} {Finken}, {and} {U.} {Seifert}: {J. Fluid Mech.} \textbf{{605}} ({2008}) {207}.

\bibitem{sui08} {Y.} {Sui}, {H.~T.} {Low}, {Y.~T.} {Chew}, {and} {P.} {Roy}: {Phys.\ Rev.\ E} \textbf{{77}} ({2008}) {016310}.

\bibitem{bagc09} {P.} {Bagchi} {and} {R.~M.} {Kalluri}: {Phys.\ Rev.\ E}  \textbf{{80}} ({2009}) {016307}.

\bibitem{kess09} {S.} {Kessler}, {R.} {Finken}, {and} {U.} {Seifert}: Eur. Phys. J. E \textbf{{29}} ({2009}) {399}.

\bibitem{lac08} {E.} {Lac} {and} {D.} {Barth{\`e}s-Biesel}: Phys.\ Fluids \textbf{{20}} ({2008}) {040801}.

\bibitem{lefe08} {Y.} {Lefebvre}, {E.} {Leclerc}, {D.} {Barth{\`e}s-Biesel}, {J.} {Walter}, {and} {F.} {Edwards-L{\'e}vy}: Phys. Fluids
  \textbf{{20}} ({2008}) {123102}.

\bibitem{nogu09a} {H.} {Noguchi}, {G.} {Gompper}, {L.} {Schmid}, {A.} {Wixforth}, {and} {T.} {Franke}: {arXiv:0811.0862 [cond-mat.soft]}.

\bibitem{jeff22} {G.~B.} {Jeffery}: Proc. R. Soc. London Ser. A \textbf{{102}} ({1922}) {161}.

\bibitem{male99} {A.} {Malevanets} {and} {R.} {Kapral}: {J.\ Chem.\ Phys.} \textbf{{110}} ({1999}) {8605}.

\bibitem{berg84} {P.} {Berg{\'e}}, {Y.} {Pomeau}, {and} {C.} {Vidal}:
  \emph{{Order within chaos: towards a deterministic approach to turbulence}} ({Wiley}, {New York}, {1984}).

\bibitem{pine05} {D.~J.} {Pine}, {J.~P.} {Gollub}, {J.~F.} {Brady}, {and}  {A.~M.} {Leshansky}: Nature \textbf{{438}} ({2005}) {997}.

\bibitem{hone92} {R.~L.} Honeycutt: Phys. Rev. A \textbf{{45}} ({1992}) {600}.


\end{thebibliography}
\end{document}